\documentclass[aps,prmaterials,preprint,longbibliography,floatfix]{revtex4-1}
\usepackage{amsmath}
\usepackage{amsfonts}
\usepackage{amssymb}
\usepackage{bm}
\def\*#1{\mathbf{#1}}
\usepackage[usenames,dvipsnames]{color} 
\usepackage{ulem}
\usepackage{graphicx}
\graphicspath{ {./images/} }
\usepackage{epstopdf}
\usepackage{xcolor}
\usepackage{enumitem} 

\usepackage{setspace}
\usepackage[english]{babel}
\usepackage[utf8]{inputenc}
\usepackage[T1]{fontenc}

\usepackage{gensymb}

\usepackage{lineno}

\usepackage{hyperref}
\hypersetup{
  colorlinks   = true, 
  urlcolor     = black, 
  linkcolor    = red, 
  citecolor   = blue 
}

\usepackage{tabularx}
\newcolumntype{Y}{>{\centering\arraybackslash}X}  
\DeclareUnicodeCharacter{2212}{-}

\begin{document}

\title{
Surface properties of alkali silicate glasses: Influence of the modifiers
}

\author{Zhen Zhang}
\affiliation{
State Key Laboratory for Mechanical Behavior of Materials, 
Xi’an Jiaotong University, Xi’an 710049, China}

\author{Simona Ispas}
\affiliation{Laboratoire Charles Coulomb (L2C), 
University of Montpellier and CNRS, F-34095 Montpellier, France}

\author{Walter Kob}
\email[Corresponding author: ]{walter.kob@umontpellier.fr}
\affiliation{Department of Physics,
University of Montpellier and CNRS, F-34095 Montpellier, France}

\begin{abstract}  
Using large-scale molecular dynamics simulations, we investigate the  surface properties of lithium, sodium, and potassium silicate glasses containing 25 mole\% of alkali oxide. The comparison of two types of surfaces, a melt-formed surface (MS) and a fracture surface (FS), demonstrates that the influence of the alkali modifier on the surface properties depends strongly on the nature of the surface. The FS exhibits a monotonic increase of modifier concentration with increasing alkali size while the MS shows a saturation of alkali concentration when going from Na to K glasses, indicating the presence of two competing mechanism that influence the properties of a MS. For the FS, we find that larger alkali ions reduce the concentration of under-coordinated Si atoms and increase the fraction of two-membered rings, implying an enhanced chemical reactivity of the surface. 
For both types of surfaces, the roughness is found to increase with alkali size, with the effect being more pronounced for the FS than for the MS. The height-height correlation functions of the surfaces show a scaling behavior that is independent of the alkali species considered: The ones for the MS are compatible with the prediction of the frozen capillary wave theory while the ones for the FS show a logarithmic growth, i.e., on the nanoscale these surfaces are not self-affine fractals.  The influence of the modifier on the surface properties are rationalized in terms of the interplay between multiple factors involving the size of the ions, bond strength, and charge balance on the surface. 
\bigskip

\end{abstract}

\maketitle

\bigskip
\section{Introduction}

Controlling the properties of the surface, such as composition, local structure, and roughness on intermediate length scales is crucial for many practical application of silicate glasses. For instance, some structural units that are uniquely present on the surface are know to control its reactivity and thus to influence the chemical stability and corrosion resistance of the surface, properties that are of great importance for biomedical and pharmaceutical usages of glasses~\cite{hench1978physical,Pantano1989,
dubois1993bonding,grabbe1995strained,hamilton1997effects,
zheng2019protein,rimola_silica_2013}. Surface morphology is another characteristic of high relevance since it has been shown to play a key role in many functional properties of glasses such as friction and adhesion~\cite{derler2009friction,belkhir2014correlation,
rahaman2015surface,li2004bacterial,biggs2002direct}. In addition, morphological investigations of glass fracture surfaces, via techniques such as atomic force microscopy (AFM), have allowed to obtain a better understanding of the failure mechanisms of glasses, documenting the importance of the surface for the mechanical properties of glasses
~\cite{gupta_nanoscale_2000,
ciccotti_stress-corrosion_2009,sarlat_frozen_2006,
ponson_two-dimensional_2006,bonamy_scaling_2006,guin_fracture_2004,
wiederhorn_use_2011,ciccotti_situ_2018,wiederhorn_roughness_2007}. 

Significant progress in our understanding of the surface properties of glasses has been made
by using a combination of experimental techniques such as 
low-energy ion scattering (LEIS) spectroscopy and AFM~\cite{kelso_comparison_1983,almeida_low-energy_2014,almeida_low-energy_2016,cushman2018low,radlein1997atomic,
poggemann2001direct,poggemann2003direct,frischat2004nanostructure}, as well as molecular dynamics (MD) computer simulations ~\cite{feuston1989topological,roder_structure_2001,
ceresoli_two-membered_2000,rarivomanantsoa_classical_2001,
du_molecular_2005,rimola_silica_2013,garofalini1990molecular,
ren2017bulk,mahadevan2020hydration,
zhang_surf-vib_2020,zhang2021abinit,zhang2021roughness}. 
Regarding the chemical composition of silicate glass surfaces, it is well established that on surfaces the concentration of oxygen and modifiers, e.g., sodium, are usually higher than in the bulk, resulting that the one of silicon atoms is lower ~\cite{kelso_comparison_1983,zhang_surf-vib_2020,zhang2021abinit}. This compositional difference between the surface and the bulk is related to their different geometry: Due to the presence of the vacuum, glass surfaces have a higher concentration of non-bridging oxygen (NBO) which in turn requires more positively charged modifiers
in their vicinity for the sake of charge  neutrality~\cite{kelso_comparison_1983}. Also the production history of the surface influences the composition and structure on the surface: 
The melt-formed surfaces (MS) of sodium silicate glasses have been found to be more enriched in Na and NBO than the fracture surfaces (FS), whereas the latter have more structural defects such as two-membered (2M) rings (i.e., closed loops of two oxygen and two silicon atoms) and under-coordinated Si~\cite{zhang_surf-vib_2020,zhang2021abinit}. This dependence has been attributed to the different mechanisms that govern the formation of these two types of surfaces: The MS  is created in the liquid state, which allows more Na to diffuse to the surface and thus to reorganize the surface structure. The FS, by contrast, is generated by creating a surface at room temperature, and therefore, one can expect that only very little reconstruction of the fracture surface has occurred once the crack has passed~\cite{zhang_surf-vib_2020}. 

A further consequence of the different production history is that for silicate glasses the FS is rougher than the one of MS and also shows a stronger dependence on composition~\cite{gupta_nanoscale_2000,zhang2021roughness}.
Moreover, experiments and simulations find that the geometrical features of the topography of the MS is described well by the theory of capillary waves that freeze in at the glass transition~\cite{gupta_nanoscale_2000,sarlat_frozen_2006,zhang2021roughness}. 
In contrast to this, early experiments indicated that the structure of the FS is given by a self-affine fractal down to the nanometer scale~\cite{pallares_roughness_2018,ponson_two-dimensional_2006}. However, this conclusion could be flawed due to the limitation in the lateral resolution of surface measurement~\cite{schmittbuhl_reliability_1995,
mazeran2005curvature,lechenault_effects_2010}. In fact, a recent large-scale MD simulation study of the FS of sodium silicate glasses with atomic resolution has demonstrated that the FS is not a self-affine fractal on length scales below 10~nm~\cite{zhang2021roughness}. However, it remains to be confirmed whether or not this conclusion also holds when the glass composition is changed, e.g., glasses containing other alkali oxides. 

Alkali oxides are a key component for modifying the properties of silicate glasses. Previous experimental and theoretical studies of alkali silicate glasses have mainly focused on their structural properties in the bulk~\cite{brawer1975raman,matson1983structure,dupree1986structure,
huang1991structural,maekawa_structural_1991,hannon1992structure,
stebbins1992structure,mysen1993structure,maehara2004structure,
uhlig1996short1,uhlig1996short2,
du2006compositional,baral2017abinitio}, and the correlations between structure and various physical properties, such as density~\cite{shartsis1952density,doweidar1996density},  phase separation and transformation~\cite{shelby1983property,kitamura2000high,uhlmann1973densification},  transport coefficients~\cite{macdonald1985low,beier1985transport,soules1981sodium}, and elastic behavior~\cite{vaills1996two,
pedone_insight_2007,zhang2022stiffness}. The observed influence of the alkali species on the various glass properties were frequently associated with the difference in cation field strength (i.e., the ratio between the charge and ionic radius of the cation) which is 1.69, 1.00 and 0.66 $e$\AA$^{-1}$ for Li, Na, and K ions, respectively~\cite{shannon1969effective}. 
It has been suggested that some anomalies of lithium silicate glasses, such as the increase of medium range order~\cite{uhlig1996short1,uhlig1996short2,du2006compositional} and elastic moduli~\cite{deguire1984dependence,pedone_insight_2007} with increasing alkali concentration, are directly related to the high field strength of lithium ions (and thus a high strength of the Li-O bond), since the other alkali silicate glasses show the opposite trend.
In contrast to this, a recent \textit{ab initio} molecular dynamics study indicates that 
the trends observed in the various physical properties of bulk alkali-silicate glasses are related to a complex interplay between multiple competing factors such as bond strength, bond length, and the local environments~\cite{baral2017abinitio}.
Also relevant for the present study is the observation that the surface tension of molten alkali silicates decreases with increasing alkali radius~\cite{shartsis1951surface} (for a given temperature and composition), which is expected to affect the properties of the glass surfaces, although so far this dependence has not been clarified.  
Thus, despite these previous studies, there are at present still many important open questions on how the chemical nature of the alkali modifiers affects the surface properties of silicate glasses, and the goal of the present work is to advance our understanding on this subject.

The main objective of this work is to investigate how the surface properties of binary alkali silicate glasses are influenced by the alkali type. Specifically, we will elucidate the effects of the modifiers on the composition, structure and topographical properties of the MS and FS. To this aim we consider the alkali species Li, Na, and K since they are among the most widely used additives for the production of commercial silicate glasses and are also the key components for chemical strengthening of glasses via ion-exchange at the glass surface~\cite{gy2008ion,varshneya2010chemical}. Our results are thus not only of practical relevance for the design of novel glasses with new/improved properties but also useful for obtaining a deeper understanding of the fracture behavior of these glasses. 

The rest of the paper is organized as follows. In Section~\ref{simulation} we introduce the simulation methods which consist of the preparation of the glass using MD and the identification of the outermost surface layer using a geometric approach. Next, in Section~\ref{results-discussion} we present and discuss the main results regarding the properties of the MS and FS. Finally, we summary and conclude this work in  Section~\ref{summary}.

\section{Simulations} \label{simulation}
\subsection{Preparation of the glass samples}
The glass samples we investigate have the nominal composition A$_2$O-3SiO$_2$ (hereafter denoted as AS3, with A = Li, Na, or K), and are produced by means of classical MD simulations. 
The interaction between the atoms are given by a pairwise
effective potential (named SHIK)~\cite{sundararaman_new_2019}, which has been found to give a good quantitative description of the density, structure and mechanical properties of bulk silica and alkali (Li, Na, K) silicate glasses~\cite{zhang_potential_2020,zhang2022stiffness}. 
Its functional form is given by
\begin{equation} 
V(r_{ij}) =  \frac{q_iq_je^2}{4\pi \epsilon_0 r_{ij}} +
A_{ij} \exp(-r_{ij}/B_{ij}) - \frac{C_{ij}}{r_{ij}^6} \quad ,
\label{eq:potential}
\end{equation}
where $r_{ij}$ is the distance between two atoms of species $i$ and $j$. In order to attain a high computational efficiency, the long-range interactions given by the Coulomb term in Eq.~(\ref{eq:potential}) are evaluated using the method proposed by Wolf {\it et al.}~\cite{wolf_exact_1999} where the long-range cutoff is set to 10~\AA; more details of this implementation can be found in Refs.~\cite{sundararaman_new_2018,sundararaman_new_2019}. A previous study~\cite{zhang_potential_2020} has shown that this treatment of the long range forces has negligible influence on various properties of the silica and sodium silicate systems while allowing for a significantly improved computational efficiency, and thus large-scale simulations can be performed.

The values of the potential parameters $A_{ij}$, $B_{ij}$, $C_{ij}$ are given in Ref.~\cite{sundararaman_new_2019}. The effective charges for Li, Na, K, and Si are 0.5727$e$, 0.6018$e$, 0.6849$e$ and 1.7755$e$, respectively. We note that these charges are compatible with those of the partial charges reported in first-principles simulations of
similar glass compositions~\cite{baral2017abinitio}. The charge for oxygen is composition-dependent in order to maintain overall charge neutrality~\cite{sundararaman_new_2019}. Our recent MD studies have shown that the SHIK potential also allows to obtain reliable results on the surface properties of silica and sodium silicate glasses~\cite{zhang_surf-vib_2020,zhang2021roughness}. The present work is the first study that uses this interaction potential for simulating the surface properties of Li and K containing silicate glasses, thus allowing to investigate how these alkali atoms modify the properties of the glass surface and how these modifications depend on the alkali type.

For each composition, approximately $2,300,000$ atoms were placed randomly in a simulation box with a volume given by the experimental value of the glass density at
room temperature~\cite{bansal_handbook_1986}. The box dimensions were roughly 20~nm, 30~nm, and 50~nm in the $x$, $y$, and $z$ directions, respectively. Previous studies have shown that such large samples are needed to allow for an accurate determination of the surface properties and the mitigation of finite size effects for the fracture process~\cite{zhang_surf-vib_2020,zhang2021roughness}. Using periodic boundary conditions in three dimensions, these samples were equilibrated at 6000~K for 80~ps in the canonical ensemble ($NVT$) and then cooled and equilibrated at a lower temperature $T_1$ (still in the liquid state) for another 160~ps. This temperature $T_1$ was 2000~K, 2000~K and 2200~K for LS3 and NS3 and KS3, respectively. (The higher $T_1$ for KS3 is to take into account the slower dynamics of the sample due to the large alkali ions.) Subsequently we cut the
sample orthogonal to the $z-$axis and inserted an empty space, thus creating two free surfaces, i.e.,~the sample has the geometry of a slab. Periodic boundary conditions were applied in all three directions. In order to
ensure that the two free surfaces do not interact with each other, the
thickness of the vacuum layer was chosen to be 14~nm. These samples were equilibrated at $T_1$ for 1.6~ns, a time
span that is sufficiently long to allow the reconstruction of the surfaces
and the equilibration of the interior of the samples. Subsequently the samples were cooled via a two-stage quenching: A cooling rate of $\gamma_1=0.125$~K/ps  was used to quench the samples from $T_1$ to a temperature $T_2$ and a faster cooling rate $\gamma_2=0.375$~K/ps to cool them from $T_2$ to 300~K. Finally, the samples were annealed at 300~K for 800~ps. The temperature $T_2$ at which the cooling rate is changed was chosen to be 1200~K which is well below the simulation glass transition temperature $T_g$ (around 1400~K). 
(Note that the determination of the $T_g$ for the sample with free surfaces is more difficult than for the bulk sample since the faster dynamics of the surface regions blurs the transition from liquid to glass.) 
At $T_2$, we also switched the simulation ensemble from $NVT$ to $NPT$ (at zero pressure) so that the generated glass samples were not under macroscopic stress at room temperature. The so obtained samples had thus two surfaces and were
used to determine the properties of the MS.

After preparation of the glass samples we introduced on one of its
free surfaces a linear ``scratch'' in the form of a triangular notch spanning
the sample in the $x$ direction of width and depth of 3~nm and 2~nm,
respectively. Subsequently we applied to the sample a strain in the $y$ direction, using a
constant rate of 0.5~ns$^{-1}$, until it broke. (Note that this rate is sufficiently low so that the fracture behavior~\cite{zhang_thesis_2020,zhang2022fracture}, as well as the properties of the resulting fracture surface do not depend on it in a significant manner~\cite{zhang_surf-vib_2020,zhang2021roughness}.) Due to the presence of the notch, the place at which the fracture initiated could be changed at will, thus allowing to reuse the undamaged glass sample for several fracture processes. 

Temperature and pressure were
controlled using a Nos\'e-Hoover thermostat and
barostat~\cite{nose_unified_1984,hoover_canonical_1985,hoover_constant-pressure_1986}.
All simulations were carried out using the Large-scale Atomic/Molecular
Massively Parallel Simulator software (LAMMPS)~\cite{plimpton_fast_1995,thompson2022lammps} with a
time step of 1.6~fs. 

The results presented in the following correspond to one melt-quench
sample for each composition. However, we emphasize that the system size
considered are sufficiently large to make sample-to-sample
fluctuations negligible~\cite{zhang_thesis_2020}. For the MS, the results for the two surfaces on
the top and bottom sides of the glass sample were averaged. For the FS, we
averaged the results over six surfaces, resulting from three
independent fracture (by changing the location of the notch). The error
bars were estimated as the standard error of the mean of the samples. 

\subsection{Identifying the surface}
In order to have a reliable description of the surface one needs a
method that allows to map the positions of the atoms onto a well-defined
mathematical surface.  The algorithm that we used for constructing this
surface mesh is based on the alpha-shape method of Edelsbrunner and
M\"ucke~\cite{edelsbrunner_three-dimensional_1994}. It starts with the
Delaunay tetrahedrization of the set of input points, i.e., the atoms in
the sample. For each of the resulting tetrahedra one evaluates its circumsphere and compares it to a reference probe
sphere that has a radius $R_\alpha$. The elements with circumsphere
radius $R$ which satisfy $R < R_\alpha$ are classified as solid, and the
union of all solid Delaunay elements defines the geometric shape of the
atomistic solid. A robust realization of this algorithm is implemented
in OVITO ~\cite{stukowski_computational_2014}.

It is important to mention that the probe sphere radius $R_\alpha$ is
the length scale which determines how many details and small features
of the solid's geometric shape are resolved. To construct the geometric
surfaces for the glass samples we use $R_\alpha = 3.2$~\AA, i.e.,
the typical distance between neighboring Si atoms. This choice allows
to resolve fine surface features but avoids the creation of artificial holes in the
constructed surfaces. We note, however, that a small change of $R_\alpha$
(e.g. by $\pm0.5$~\AA) will not alter significantly the results presented in
the the following (see~Refs.~\cite{zhang_thesis_2020,zhang_surf-vib_2020}
for details). Note that, by definition, the surfaces are two-dimensional objects. Finally, we point out that for the FS  we have not considered the parts of the surface that are closer than $\approx5$~nm to the
top/bottom MS in order to avoid the influence of these surfaces onto
the properties of the FS.

Once the geometric surface is constructed, i.e., the mesh points of
the surface are identified, we first fit a plane to the set of mesh
points using a least squares fitting procedure, and defined this plane to be 
$z=0$. Finally, a linear interpolation is applied to the triangular
mesh to obtain a uniform quadratic grid which is subsequently used to
determine the morphology and roughness of the surface. 
Recent studies that used similar simulation protocols and surface analysis strategies have demonstrated that this approach does indeed allow to determine the surface characteristics of silica and sodium silicate glasses~\cite{zhang_thesis_2020,zhang_surf-vib_2020,zhang2021roughness}.

\section{Results and discussion} 
\label{results-discussion}

\bigskip

\begin{figure}[ht]
\center
\includegraphics[width=\textwidth]{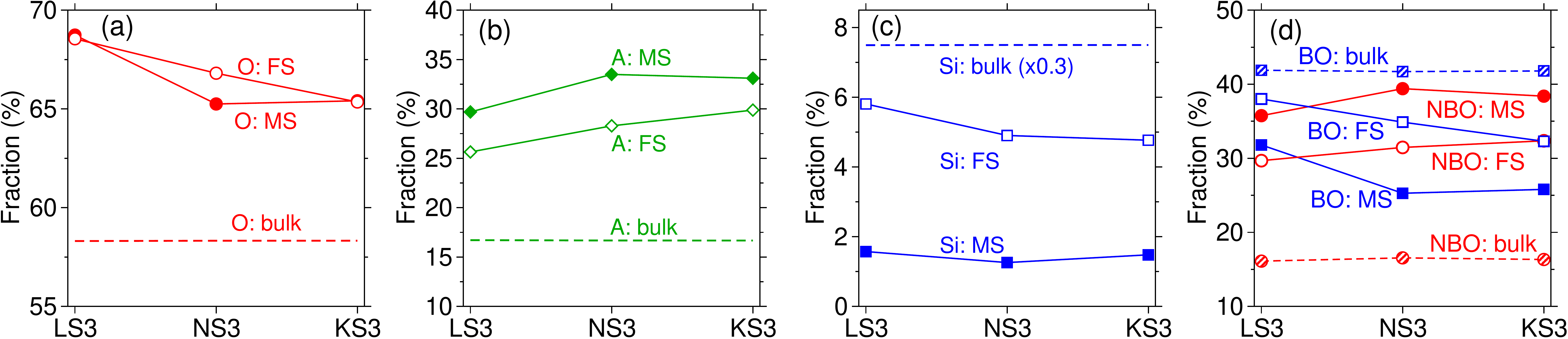}
\includegraphics[width=\textwidth]{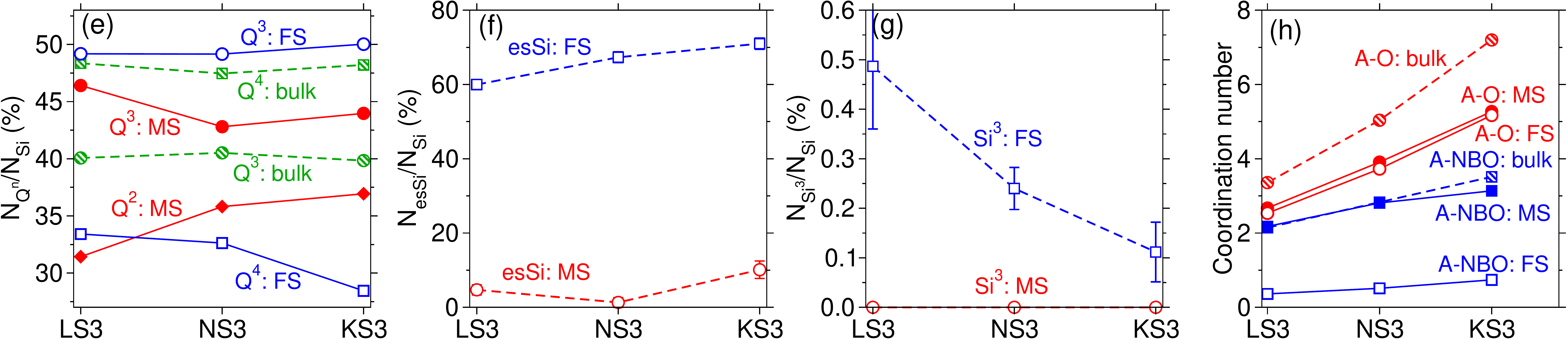}
\caption{
Surface composition and structure. (a)-(c) Elemental concentration of O, alkali (A), and Si, respectively, for the two types of surfaces and the bulk (dashed line). (d) Fraction of bridging and non-bridging oxygen atoms, BO and NBO, respectively. The fractions are with respect to the total number of surface (bulk) atoms. (e) Fraction of the  most abundant types of $Q^n$ species on the surfaces. (f)-(g) Fraction of edge-sharing Si (esSi) and three-fold coordinated Si (Si$^3$), respectively. Note that in panels (e)-(g) the fractions are relative to the total number of Si on the surface. Error bars smaller than the symbol size are not shown. (h) Coordination number of the modifiers. The cutoff distances for determining the coordination number of the alkali are 2.5, 3.0, and 3.6~\AA\ for Li, Na, and K, respectively, corresponding to the locations of the first minimum in $g_{\rm AO}(r)$.\\
}
\label{fig_comp-struc-as3}
\end{figure}

\subsection{Composition and structure of the surface}
In this subsection we present how the composition and local structure of the surface depends on the alkali type and how the presence of the surface affects the structural properties of the sub-surface layer.

Figure~\ref{fig_comp-struc-as3} shows how the surface composition and the major structural quantities depend on alkali and surface type. Overall, we note that the influence of the modifiers are more complex than the simple monotonic dependence that one might expect, in that several quantities saturate or show even a non-monotonic behavior as a function of the alkali type, indicating the presence of competing mechanisms. The differences between the two types of surfaces can be attributed to their production history which affects the thermodynamics and kinetics of their formation (see below for a more detailed discussion). 
Panels (a) and (b) show that, with respect to the bulk, both kinds of surfaces are significantly enriched in alkali and oxygen and that this enrichment in alkali is more pronounced for the MS, results that agree with a previous simulation study on sodium silicate glass surfaces~\cite{zhang_surf-vib_2020}. Consequently the two surfaces are strongly depleted of Si relative to the bulk, panel (c), and this depletion is more pronounced for the MS. The compositional difference between the MS and the FS can be related to the fact that the MS was created in the liquid state, which allows more alkali ions to diffuse to the surface, resulting in a more pronounced change of composition with respect to the bulk. In contrast to this, the FS was generated by dynamic fracture (i.e., with a crack velocity on the order of 10$^3$~m/s)~\cite{zhang_thesis_2020, zhang2022fracture} at room temperature, and thus the atomic diffusion is very limited, resulting that the compositional change at the FS is smaller than the one of the MS~\cite{zhang_surf-vib_2020}. Furthermore, one notices that the compositional difference depends not only on the surface type but also on alkali species: With increasing alkali size, the composition of the FS changes monotonically, while the MS shows a notable change of surface composition from LS3 to NS3, but basically no difference between NS3 and KS3, implying a possible saturation of the modifier effect. 
The monotonic change of composition from LS3 to KS3 for the FS can be rationalized by considering the difference in the A-O bond strength: The force constants inferred from the frequency of the characteristic vibrational modes of the A-NBO interaction have been reported to be 347~cm$^{-1}$, 180~cm$^{-1}$, and 113~cm$^{-1}$ for Li-O, Na-O, and K-O, respectively~\cite{hauret1995dynamic}. Therefore, Li atoms are bound more strongly to the matrix than the two other alkali atoms and hence their thermodynamic driving force for migration to the surface is smaller.  For the FS, the saturation of composition from NS3 to KS3 might be attributed to the delicate balance between two trends, i.e., the aforementioned bond strength and ionic size, with the latter resulting in a decreasing mobility of the large alkali ions.

The deviation of the surface composition from the bulk has the consequence that the net charge on the surface is not zero. We find that the MS as well as the FS are negatively charged and the negativity (per atom) is more pronounced for the MS than for the FS. This imbalance of surface charge weakens with increasing alkali size since larger alkali ions have a larger positive effective charge~\cite{sundararaman_new_2019}. We also mention that this charge imbalance decreases quickly with distance from the surface and thus charge neutrality is restored within a few nm below the surface, see below for details.

Next, we show that not only the surface composition, but also the network connectivity is influenced by the alkali type. This is reflected in the fraction of bridging oxygen (BO), i.e., oxygen that is bonded to two Si, and thefraction of non-bridging oxygen (NBO), i.e., oxygen that is bonded to a single Si and thus forms a dangling SiO$^-$ bond, Fig.~\ref{fig_comp-struc-as3}d. Firstly, we note that in the bulk the alkali type does not affect the network connectivity, in agreement with previous findings~\cite{pedone_insight_2007, zhang2022stiffness}. Secondly, one observes that both the MS and the FS are significantly more depolymerized than the bulk structure and the depolymerization is more prominent for the MS (i.e., higher fraction of NBO). This result is directly related to the fact that the MS is more enriched in modifiers, panel (b), and one sees that these two quantities do indeed track each other very closely. In fact, one finds that the ratio NBO/alkali is around 1.0 (bulk), 1.2 (MS), and 1.1 (FS) and depends only weakly on alkali type. 

\begin{figure}[t]
\center
    \includegraphics[width=0.8\textwidth]{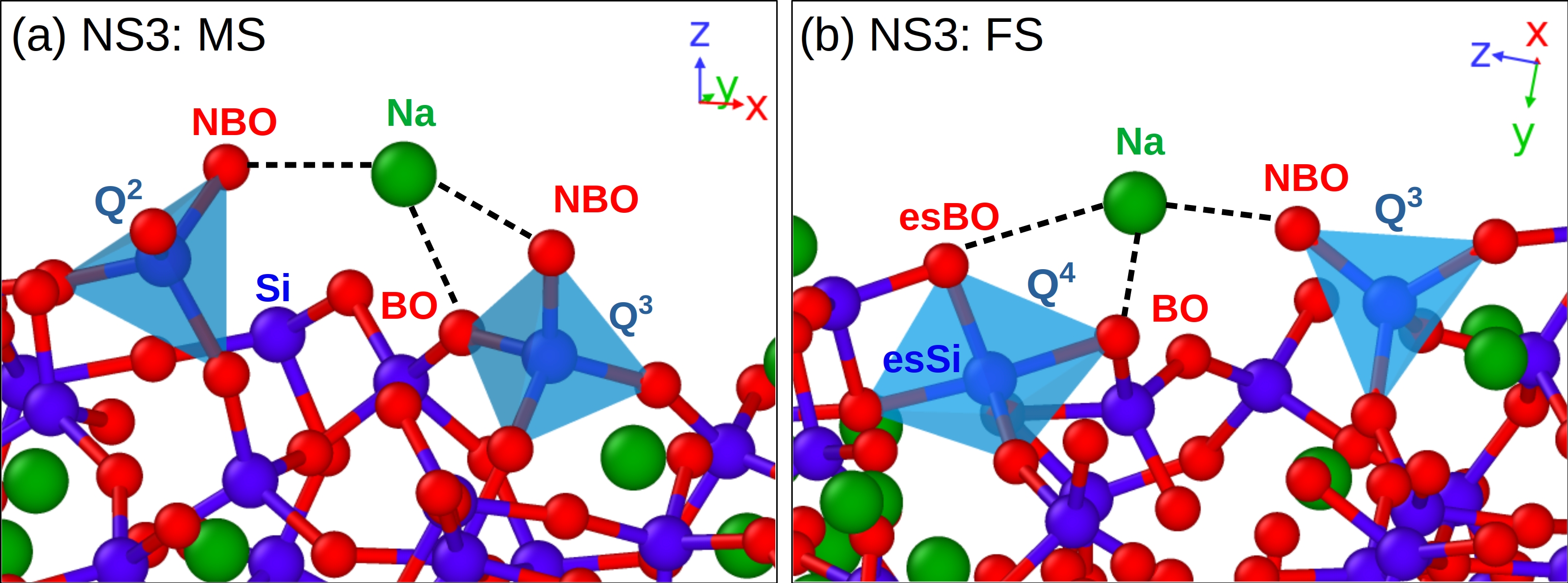}
    \caption{Snapshots of the representative atomic configuration near the MS (a) and the FS (b) of the NS3 glass. O, Si and Na are represented by spheres in red, blue and green, respectively. The Si-O bonds that are shorter than 0.2~nm are denoted by solid sticks. The interactions between a Na and its nearest neighbor oxygens are indicated by dashed lines. The two [SiO$_4$] tetrahedra that are interacting with the Na are shaded light blue and are labeled by their $Q^n$ type. Bridging oxygen (BO), non-bridging oxygen (NBO), and edge-sharing (es) atoms in the neighborhood of the Na are also labeled.
    }
    \label{fig_surf-struc-snapshot}
\end{figure}

To characterize the network connectivity on length scales larger than the nearest neighbor distance, one can probe the concentration of the so-called  $Q^n$ units, where $n$ denotes the number of BO in a [SiO$_4$] tetrahedron. Panel (e) shows the fraction of the two most abundant $Q^n$ units for the surfaces as well as the bulk and one recognizes that the bulk sample is mainly composed of $Q^4$ and $Q^3$ units, demonstrating that the Si-O network is highly cross-linked, independent of the alkali type considered. These $Q^n$ values for the bulk are in good agreement with the ones reported in a previous simulation study of the structure of alkali silicate glasses~\cite{du2006compositional}, but are about 20\% off the values given by NMR~\cite{maekawa_structural_1991}. This discrepancy is likely a consequence of the high cooling rates typically used in MD simulations which  exceed by orders of magnitude the ones of experiments~\cite{vollmayr_cooling-rate_1996,li_cooling_2017}. 
In comparison with the bulk, the surface structures are significantly less connected, favoring the formation of less-polymerized 
units in that the MS has no longer $Q^4$ units but is instead mainly composed of $Q^3$ and $Q^2$ structures  (the latter are presents in the bulk only with a small concentration) and for the FS one finds more $Q^3$ units than $Q^4$, opposite to the case in the bulk (see also the snapshots in Fig.~\ref{fig_surf-struc-snapshot}).

Figures~\ref{fig_comp-struc-as3}(f) and (g) demonstrate that the nature of the modifier influences also the concentration of the defect structures in the network and that this
effect is particularly pronounced for the FS since this surface was generated from a highly non-equilibrium dynamic fracture process. Panel (f) shows that over 60\% of Si atoms on the FS are edge-sharing (esSi), i.e., forming two-membered (2M) ring structures (see the snapshot in Fig.~\ref{fig_surf-struc-snapshot}b), and that this probability grows with increasing alkali size. (Note that basically no 2M rings are found in the bulk since these structures are energetically very unfavorable.) This finding can be linked to the dependence of the medium range order on alkali type: A recent simulation study, using the same potential as in the present work, has demonstrated that in (bulk) alkali silicate glasses with 20\% alkali oxide, the fraction of small sized rings (less than 5 Si-O linkages) increases as the alkali varies from Li to K~\cite{sundararaman_new_2019}. These small rings, which are likely to be broken during fracture because of their high internal stress, are favourable structures for the formation of 2M rings since the creation of
larger rings, although energetically more favorable than short rings, takes much more time because it requires an atomic rearrangement on larger length scale~\cite{zhang_surf-vib_2020}. 
Panel (f) also shows that for the MS the concentration of 2M rings is very small and, panel (g), that also the fraction of Si$^3$ is basically zero (for the FS the latter concentration is significantly higher). These low concentrations can be attributed to the fact that for the MS the high temperature at its creation as well as the long time for the cooling allows the surface structure to reorganize and thus to eliminate these energetically unfavorable structural units. 

Panel (g) shows that for the FS the fraction of three-fold under-coordinated Si, Si$^3$, although small in absolute value, is reduced by a factor of five when increasing the alkali size from Li to K. This strong dependence is likely due to the above mentioned fact that the alkali-NBO energy decreases significantly with increasing alkali radius, thus making it easier for a under-coordinated Si atom to pick up an additional oxygen atom. However, it cannot be excluded that some other aspects of the fracture process or the relaxation of the local structure shortly after the fracture depend on the alkali species, thus giving rise to the observed dependence of the Si$^3$ concentration.

Finally, we demonstrate that also the local environment of the modifiers on the surface is different from the one in the bulk and that it depends on the alkali type. To quantify this dependence, we have determined the coordination numbers, $Z_{\rm A-O}$ and $Z_{\rm A-NBO}$, of the modifiers by counting the number of oxygen and NBO atoms, respectively, up to the first minimum in the pair distribution function of A-O, $g_{\rm AO}(r)$. Panel (h) shows that for the bulk the mean $Z_{\rm A-O}$ are respectively 3.4, 5.0, 7.2, for Li, Na, and K, comparable with the values reported in previous simulation works for the same glass compositions but using different interaction potentials~\cite{du2006compositional, pedone_insight_2007}.
The coordination numbers of the alkali ions on the surfaces are smaller relative to the bulk values: For the MS, the mean $Z_{\rm A-O}$ are 2.7, 3.9, and 5.3 for Li, Na, and K, respectively and the corresponding values for the FS are only slightly smaller. This considerably reduced coordination number of the modifiers on the surface is due to the fact that the modifiers are located at the outermost atomic layer and thus have no oxygen atoms on the vacuum side to be bonded to. 
For the MS the coordination number between a modifier and a NBO is basically the same as in the bulk, i.e, the reduction of the coordination number found in $Z_{\rm A-O}$ is not seen. This implies that the decrease in $Z_{\rm A-O}$  at the surface is because the modifier reduces the number bridging oxygen atoms but not NBO, since the latter can be more easily accommodated at the surface, in agreement with panels (d) and (e).
The value of $Z_{\rm A-NBO}$ is around 2.5 and slightly increasing with increasing alkali size. This value
can be related to the fact that the most abundant $Q^n$ units on the MS are $Q^3$ and $Q^2$, contributing respectively one and two NBOs for potential bonding with the alkali on the surface, see panel (e). By contrast, the FS is primarily composed of $Q^3$ and $Q^4$ and consequently the number of candidate NBOs for bonding with the modifier is significantly less, resulting in a considerably smaller mean coordination number $Z_{\rm A-NBO}$ around 0.5, a value that increases slightly with increasing alkali size. 
The snapshots of the two surfaces of the NS3 glass, Fig.~\ref{fig_surf-struc-snapshot}, illustrate the representative atomic structures and coordination environments near the surfaces. The bulk (although it is more polymerized than the surfaces) has a $Z_{\rm A-NBO}$ comparable to that of the MS which can be ascribed to the presence of NBOs in all directions around the modifier, contrasting the situation on a surface. Finally, we note that the modifier effect is reflected by the fact that the $Z_{\rm A-NBO}$ generally increases with increasing alkali size.

\begin{figure}[t]
\center
    \includegraphics[width=0.95\textwidth]{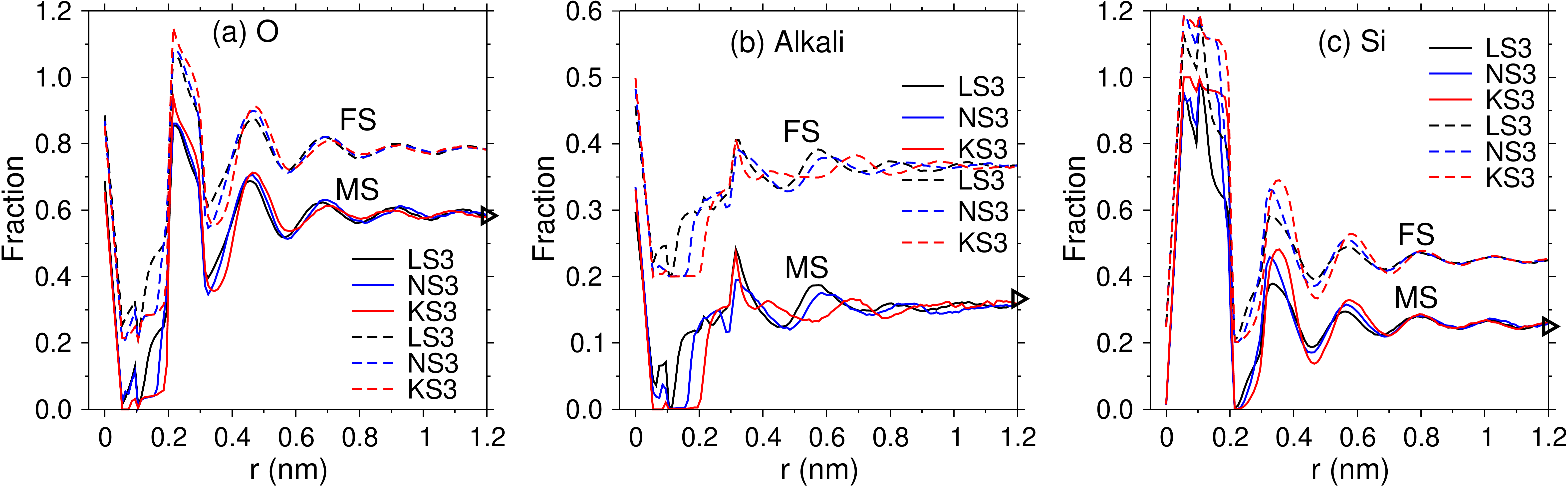}
    \includegraphics[width=0.95\textwidth]{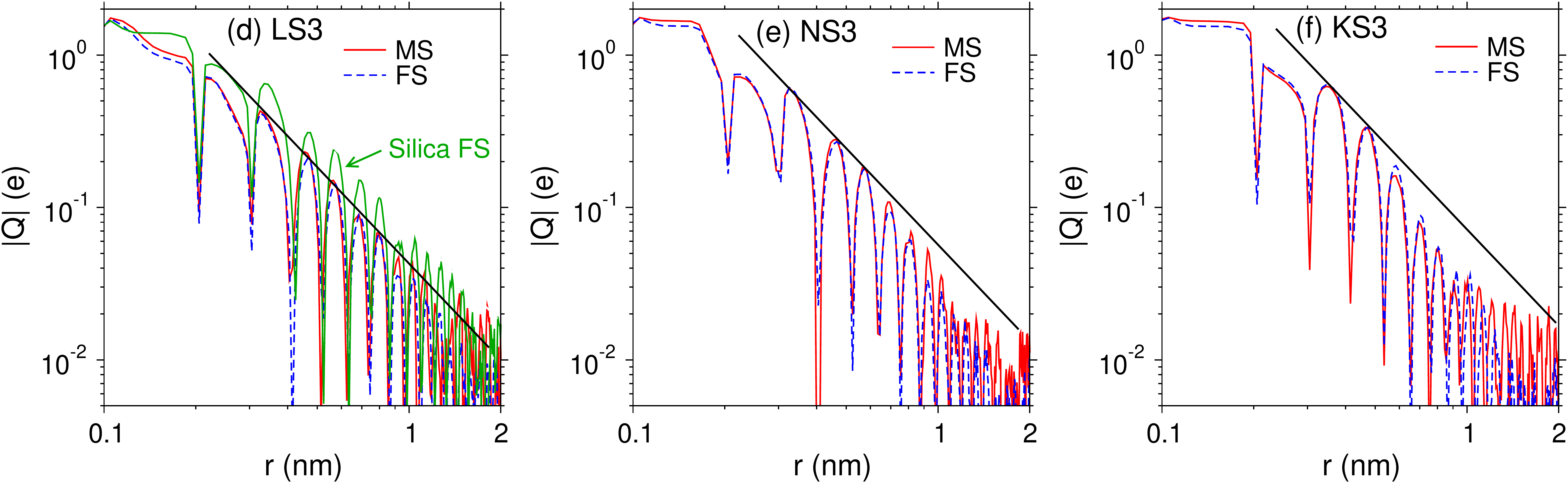}
\caption{(a)-(c) Depth profiles of elemental concentrations with respect to the surfaces. For better visibility the data for the FS are shifted upward by 0.2. In practice, the composition at distance $r$ is the mean over a layer with thickness 0.11~nm. The triangles on the right ordinate axis 
indicate the corresponding bulk values. (d)-(f) Double-logarithmic plots of the absolute value of the atomic charge $Q$ as a function of distance from the surface. 
The black solid lines are power-laws with exponent -2 that help to distinguish the decay behavior of $Q$. For the sake of comparison we include in (d) also the data for the FS of silica glass~\cite{zhang_surf-vib_2020}.\\
} 
    \label{fig_depth-profile}
\end{figure}
In addition to the properties of the surface monolayer, we have also investigated how the composition evolves with increasing distance $r$ from the surface.  Here we define this distance as the length of the shortest path from a given atom below the surface to any atom on the surface, and thus $r = 0$ represents the surface monolayer. Figures~\ref{fig_depth-profile}(a)-(c) show that for $r \leq 0.1$~nm, the fractions of O and Na decrease very rapidly with $r$, while the Si concentration increases steeply, indicating that the top-most surface layer is strongly enriched in O and alkali and is followed by a layer that is mainly composed of Si atoms. This layered arrangement of different atomic species propagates towards the interior of the glass, although the fluctuations of the composition gradually decay with increasing $r$. 
For the Si concentration the details of this layering effect are found to depend on the type of alkali atom in that larger modifiers give rise to more pronounced peaks in the region close to the surface, i.e., less than 0.7~nm, see panel (c), and at the same time the location of the peaks shifts to slightly larger distances. These effects can be directly ascribed to the size of the alkali ions since their presence at the surface will push the Si atoms to larger $r$. From panel (b), one observes that the fluctuation of the alkali concentration seems to decay faster than that of the Si and O, reaching the bulk value (indicated by the triangles on the right ordinate) already at $r \approx 1$~nm. This observation can be attributed to the higher mobility and thus higher flexibility of the alkali ions relative to the Si and O~\cite{zhang_surf-vib_2020}. 

\begin{figure}[t]
\center
    \includegraphics[width=0.95\textwidth]{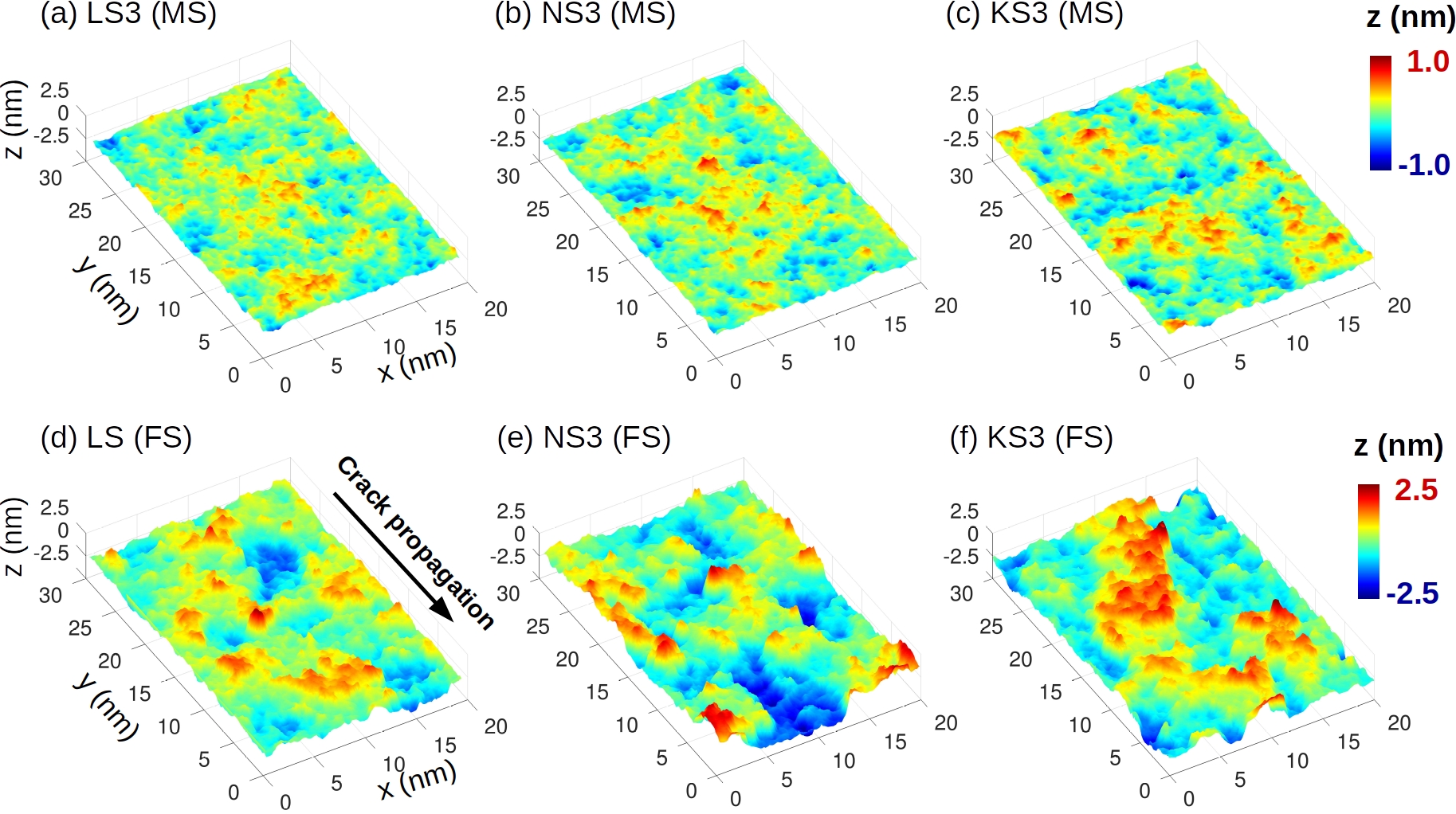}
    \caption{Topography of the surface. Melt-formed surfaces (a)–(c) and fracture surfaces (d)–(f) for the AS3 glasses. For the FS, the crack propagates in the negative $y$-direction and the crack front is parallel to the $x$-direction. The mean surface height is equal to zero. \\
}
    \label{fig_surf-images}
\end{figure}

To quantify the decay of the compositional fluctuation with increasing distance from the surface, we use the per-atom atomic charge as a overall structural indicator, defined by  $Q(r)=\sum_\alpha  q_\alpha f_\alpha(r)$, where $q_\alpha$ and $f_\alpha$ are, respectively, the charge and fraction of atomic species $\alpha$. Figures~\ref{fig_depth-profile} (d)-(f) show the $r$-dependence of the absolute value of the charge $Q$ in a double-logarithmic representation. For LS3, one recognizes that the decay of $Q$ is described well by a power-law with an exponent -2. Surprisingly we find that this $r$-dependence as well as the value of the exponent are very similar to those found for the behavior of $Q(r) $ in pure silica~\cite{zhang_surf-vib_2020}, included in the graph as well. Our results are thus coherent with the view that Li acts as a pseudo-network former which enhances the cohesion of the Si-O network structure~\cite{pedone_insight_2007}. In contrast to LS3, the decay of $Q$ with $r$ for NS3 and KS3 can no longer be described by a power-law but is rather exponential-like with a decay rate that is larger for KS3 than for NS3, panels (e) and (f). These results can be attributed to the fact that silica and LS3 can be considered as structures that are relatively homogeneous and therefore the elastic response of the medium due to a defect (here the surface) decays algebraically with distance. In contrast to this, Na and K are strongly modifying the network resulting in pronounced structural heterogeneities, making that the perturbation by the surface is damped out much quicker, i.e., exponential-like.
 
\subsection{Topography of the surface}
In this subsection we discuss the structure of the surface on large scales and probe how it depends on the alkali type and the preparation protocol. 

To get a first idea on the topographical features of the surfaces we present in Fig.~\ref{fig_surf-images} a map of the height of the MS and FS, top and bottom panels, respectively. (The $z = 0$ level has been determined such that the average height is zero). Overall, we note that the MS are smoother than the FS and that the roughness of the MS seems to be independent of the alkali type, while the one for the FS increases with alkali size. In order to makes these statements more precise we will in the following quantify the roughness and the symmetric property of the surfaces.

\begin{figure}[t]
\center
\includegraphics[width=0.7\textwidth]{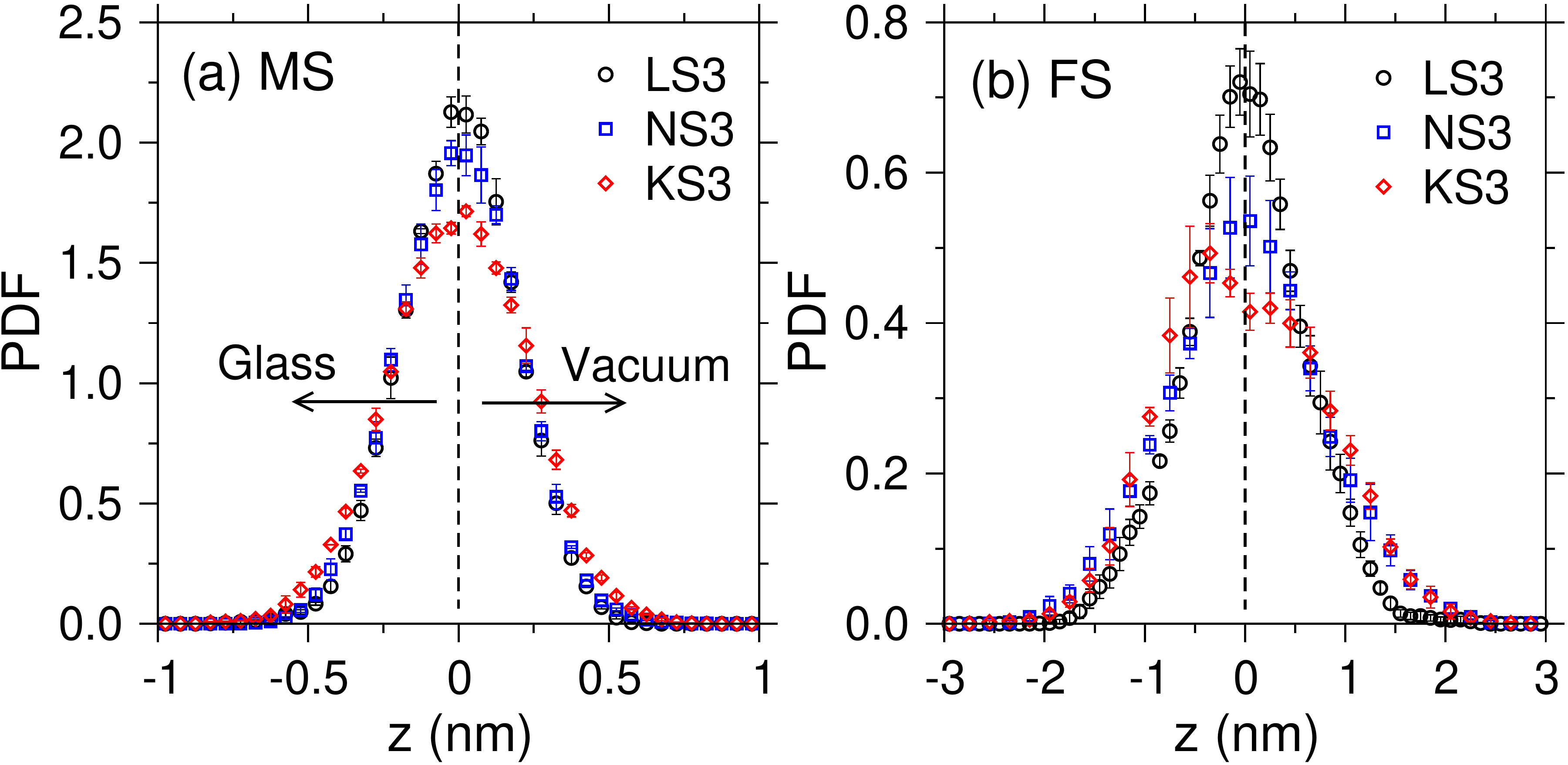}
\includegraphics[width=0.7\textwidth]{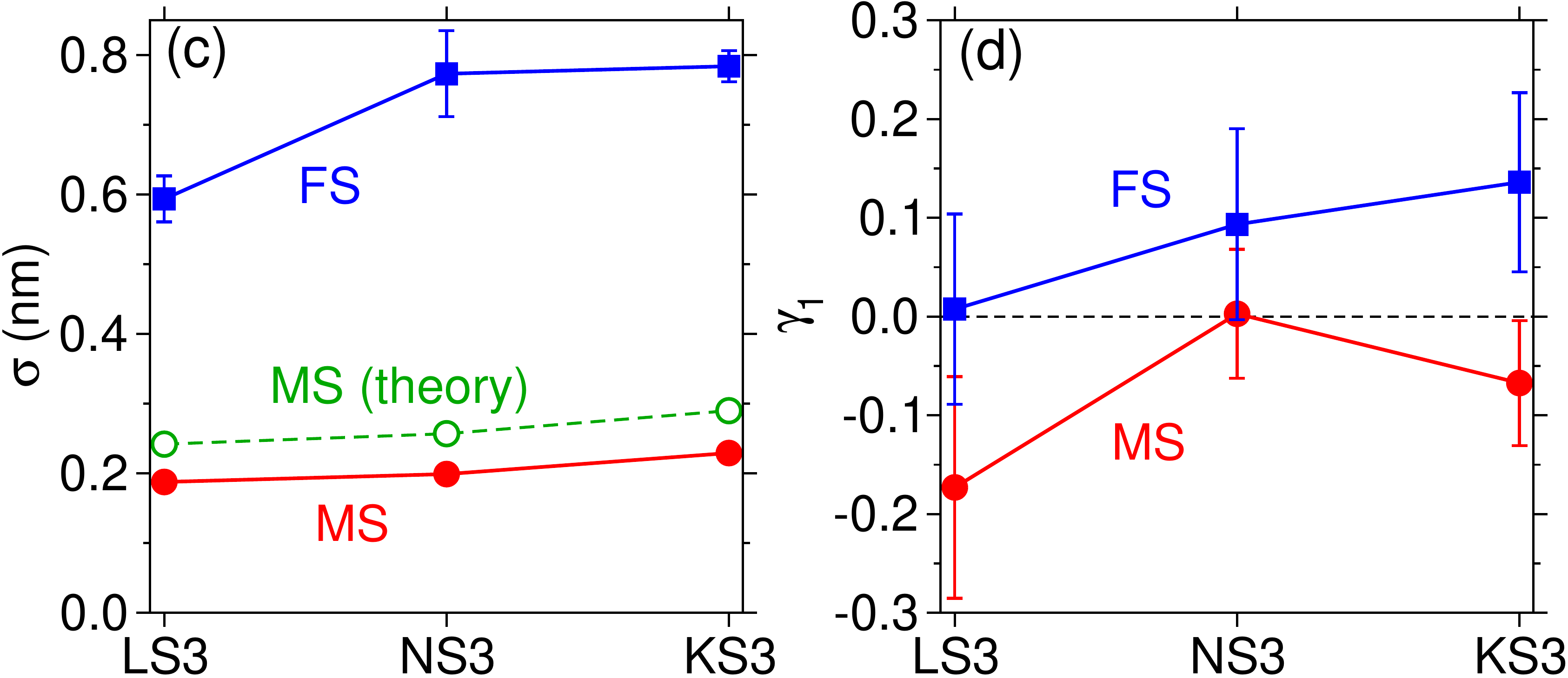}
\caption{ (a-b) Distribution of surface height for MS and FS, respectively. The data are shown for surfaces with an area of 20~nm $\times$ 30~nm, corresponding to Fig.~\ref{fig_surf-images}. The mean surface height is equal to zero. (c-d) Compositional dependence of the roughness and the skewness $\gamma_1$ of the surface height distribution, respectively. The dashed line with open symbols is a theoretical prediction of the intrinsic roughness of melt-quenched surfaces. For the sake of comparison, we note that the roughness for the MS and FS of a silica glass is 0.25~nm and 0.42~nm, respectively~\cite{zhang2021roughness}.
    }
    \label{fig_surf-roughness}
\end{figure}

Figures~\ref{fig_surf-roughness} shows the distribution of the surface height $z$ for the three glasses (MS and FS, panels (a) and (b), respectively). We recognize that these distributions have a Gaussian-like shape and that the one for the MS depends only weakly on the alkali type while for the FS this dependence is more pronounced. This observation reflects thus again the fact that the MS has a structure that has been well equilibrated while the one of the FS is strongly affected by dynamical processes that do not allow the structure to relax.

The standard deviation of the height distribution defines the roughness $\sigma$ of the surface. Panel (c) shows that for the MS this roughness has a value of $\approx$\;0.2~nm and increases weakly with increasing alkali size. This result can be rationalized by recalling the concept of a frozen liquid interface~\cite{jackle_intrinsic_1995,seydel_freezing_2002}, which predicts that the intrinsic roughness of a MS is determined by the freezing of the capillary waves at a temperature $T_0$ during cooling, i.e., $\sigma \approx \sqrt{k_{\rm B}T_0/ \gamma}$, where $k_{\rm B}$ is Boltzmann's constant and $\gamma$ is the surface tension at $T_0$. Previous studies have shown that this theory works well if one uses for $T_0$ the glass transition temperature $T_g$~\cite{gupta_nanoscale_2000,roberts2005ultimate,sarlat_frozen_2006,zhang2021roughness}.   
Making the assumption that $T_g=1400$~K is independent of glass composition and using the surface tension data of the corresponding alkali silicate melts ~\cite{shartsis_surface_1951} ($\gamma$ is 0.312, 0.278, and 0.218 J/m$^2$ for LS3, NS3, and KS3, respectively), allows to predict how the intrinsic roughness of the MS depends on the alkali species. Panel (c) shows that this theoretical prediction (open symbols) follows the same compositional dependence as the simulation results, although the values are about 20\% higher. This overestimation is likely related to the fact that the surface can relax to some extent even for temperatures below $T_g$, thus decreasing its roughness. On overall the good agreement between the predicted and measured values of $\sigma$ gives thus support to the view that surface tension is the controlling factor in determining the roughness of MS. Also of interest is the fact that these values of the roughness are smaller than the one found for the MS of silica, which is $\sigma$=0.25~\cite{zhang2021roughness}. This shows that the presence of the alkali atoms leads to a flattening of the surface because of the enhanced mobility of these ions with respect to the one of the network.

The roughness of the FS is more than three times higher than the one of the MS, demonstrating that the formation process has a significant impact on its topography. The roughness increases by about 20\% when  lithium is replaced by sodium, and below we discuss this trend in more detail. If sodium is replaced by potassium, the amplitude of these fluctuations does, however, not increase further, in line with the remarks made in context of Fig.~\ref{fig_comp-struc-as3}(b) that for Na and K the increased mobility due to the  decreasing bonding strength is compensated by an increasing ion size that hinders the migration of the atoms and as a result the geometrical properties of the surface do not change significantly. We are not aware of any experimental data for the surface roughness of AS3 glasses and therefore cannot compare our predictions with real data. However, for the case of silica such a comparison has shown that the prediction of the simulation is accurate to within 10\%~\cite{zhang2021roughness}, giving credibility to the results presented here.

The trend that the roughness increases with alkali size can be related to the fact that with increasing alkali size the bonds with the network become weaker, thus allowing for larger fluctuations in the surface height. A further mechanism that enhances this dependence is related to the observation that the glass becomes increasingly ductile when changing from Li-silicate to K-silicate glasses~\cite{baral2017abinitio}. 
This increased ductility originates from the enhanced structural heterogeneities, particularly the medium range order~\cite{sundararaman_new_2019}: With increasing alkali size, the ring size distribution becomes broader (the frequency of the small- as well as large-sized rings increases while the one for intermediate-sized rings decreases). This leads to a larger variety of structures along the fracture path and thus a fracture surface that is rougher. This trend is also in line with the fact that for silica, which has a relatively narrow ring size distribution, the roughness of the FS is found to be around 0.42~nm~\cite{zhang2021roughness}, thus a value that is significantly smaller than the ones found for the AS3 glasses.

A further question of interest is whether or not the two sides of the surface (facing the vacuum or facing the glass) are statistically equivalent. For this, we have investigated the symmetry of
the surfaces by quantifying the skewness, $\gamma_1 = \langle z^3\rangle/(\langle z^2\rangle)^{3/2}$, 
of the surface height
distribution. 
Figure~\ref{fig_surf-roughness}(d) shows that both surfaces are asymmetric: The MS are more likely to have deep holes than high protrusions on the surface side facing the vacuum ($\gamma_1$ is negative), whereas the FS shows the opposite, which is directly related to the fact that during the fracture process the breaking of Si-O-Si bridges or chain-like structures gives rise to
a spiky surface~\cite{zhang2021roughness}. The influence of the modifier is clearly seen for the FS in that larger alkali ions give rise to more asymmetric surface, whereas the influence of alkali type on the symmetry property of the MS is not evident. Finally we note that a non-vanishing $\gamma_1$ indicates that the capillary wave theory cannot be valid in a strict sense since this approach predicts $\gamma_1=0$~\cite{zhang2021roughness}.

To characterize the topography of the surfaces on larger length scales it is useful to look at the height-height correlations. For this we define the normalized surface height autocorrelation function 

\begin{equation}
C(r)=\dfrac{\langle z(r_0)\cdot z(r_0+r) \rangle - \langle z(r_0) \rangle \langle z(r_0+r) \rangle}{\langle z^2(r_0) \rangle - \langle z(r_0) \rangle^2} \quad ,
\end{equation}

\noindent
where $z(r_0)$ is the surface height at a reference position $r_0$, and $z(r_0+r)$ is the height at a distance $r$ from the reference point. The brackets $\langle ... \rangle$ denote the average over the position $r_0$ and the different realizations of the surfaces. Since by construction of the surface we have that $\langle z(r_0) \rangle=0$ and $\langle z^2(r_0) \rangle - \langle z(r_0) \rangle^2$ is equal to $\sigma^2$, the above expression can be simplified to

\begin{equation} \label{eq_Cr}
C(r)=\dfrac{\langle z(r_0)\cdot z(r_0+r) \rangle}{\sigma^2} \quad .
\end{equation}

\begin{figure}[ht]
\center
    \includegraphics[width=0.9\textwidth]{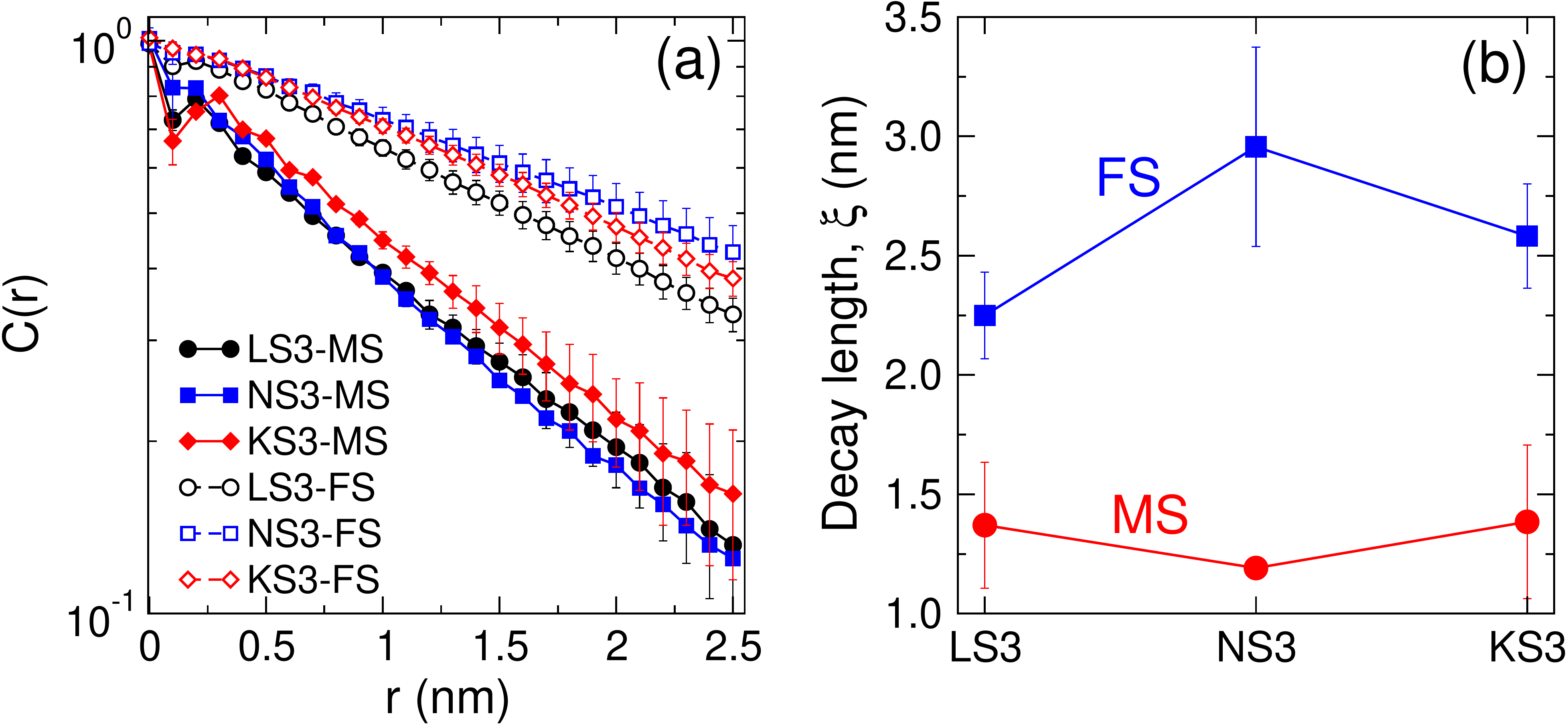}
    \caption{(a) Normalized surface height autocorrelation function $C(r)$, see Eq.~(\ref{eq_Cr}), for the AS3 glass surfaces. (b) The decay length $\xi$ estimated by fitting $C(r)$ by the expression $C(r)=A{\rm exp}(-r/\xi)$.\\
    }
    \label{fig_surf-acf}
\end{figure}

\noindent (Note that the MS is statistically isotropic and hence $C(r)$ depends only on the distance $r$. For the FS this is not quite true, since the propagation direction of the crack front gives rise to an anisotropy. It is found, however, that this effect is relatively mild~\cite{zhang2021roughness}, and hence here we have averaged over all directions.) 
From Fig.~\ref{fig_surf-acf}(a) one recognizes that for all investigated cases $C(r)$ exhibits an exponential decay and that for the FS this decay is slower than the one for the MS. We determine the decay length $\xi$ by fitting $C(r)$ in the range $0.5 \leq r \leq 1.8$~nm by the functional form  $C(r)=A\,{\rm exp}(-r/\xi)$. Figure~\ref{fig_surf-acf}(b) shows that $\xi_{\rm MS}$ and $\xi_{\rm FS}$ have a value of $\approx \!1.3$~nm and $\approx\!2.5$~nm, respectively, and depend only mildly on the alkali type.

\begin{figure}[ht]
\center
\includegraphics[width=0.9\textwidth]{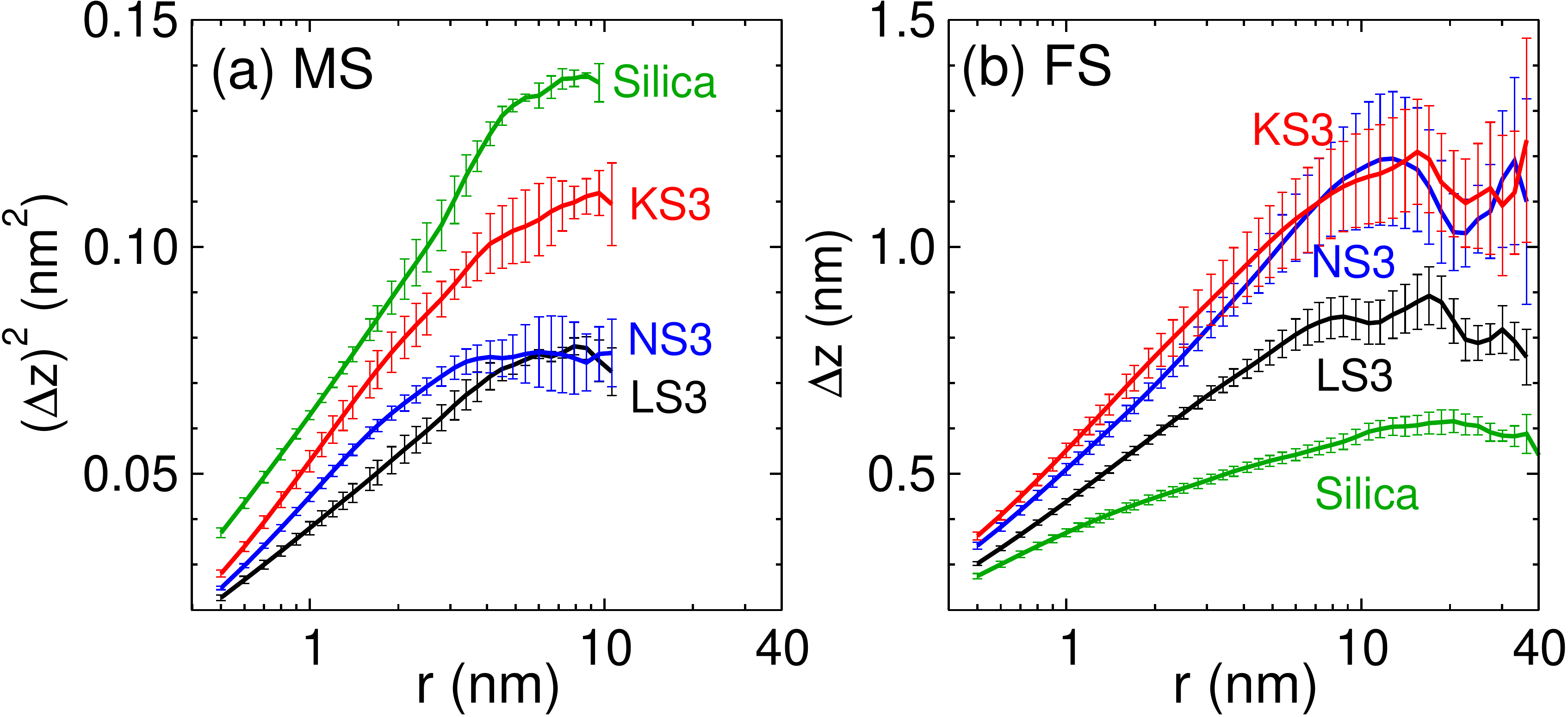}
\caption{
Height-height correlation function $\Delta z(r)$, see Eq.~(\ref{eq:Deltaz}), (linear-log scale) for
the MS (a) and FS (b). In (b) the curves are for the direction of crack propagation. The data for a silica glass~\cite{zhang2021roughness} are included for comparison.
    }
    \label{fig_surf-scaling}
\end{figure}

A further possibility to characterize the topography of the surface is to consider the one-dimensional height-height correlation function 

\begin{equation}
\Delta z(r) = \sqrt{\big \langle [z(r+x) - z(x)]^2 \big \rangle_x} \quad ,
\label{eq:Deltaz}
\end{equation}

\noindent
which measures the variance of the height difference between two points separated by a distance $r$
along a direction $x$ \cite{zhang2021roughness,bares_nominally_2014,wiederhorn_roughness_2007,pallares_roughness_2018,vink2005capillary2005}.
Figure~\ref{fig_surf-scaling}(a) shows $(\Delta z)^2$ as a
function of $r$ for the MS and, for the sake of comparison, we have included in this graph also the data for the MS of silica glass~\cite{zhang2021roughness}. 
One sees that, for length scales less than $\approx\;$4~nm,  $\Delta z(r)$ increases logarithmically with $r$, in accordance with the prediction of the frozen capillary wave theory, i.e., $(\Delta z)^2 \propto$~ln~$r$~\cite{jackle_intrinsic_1995,seydel_freezing_2002}. For larger length scales the curves become flat, most likely because of finite size effects which prevent the fluctuations to grow beyond a level that is dictated by the linear extension of the sample. For a fixed distance $r$ the value of $\Delta z$ increases with increasing alkali size and one also notices that the curve for silica is the highest one. This dependence of $\Delta z(r)$ on the composition is thus in harmony with the one of the roughness $\sigma$ presented in Fig.~\ref{fig_surf-roughness}(c) and which shows the same ordering.

\begin{figure}[th]
\center
    \includegraphics[width=0.9\textwidth]{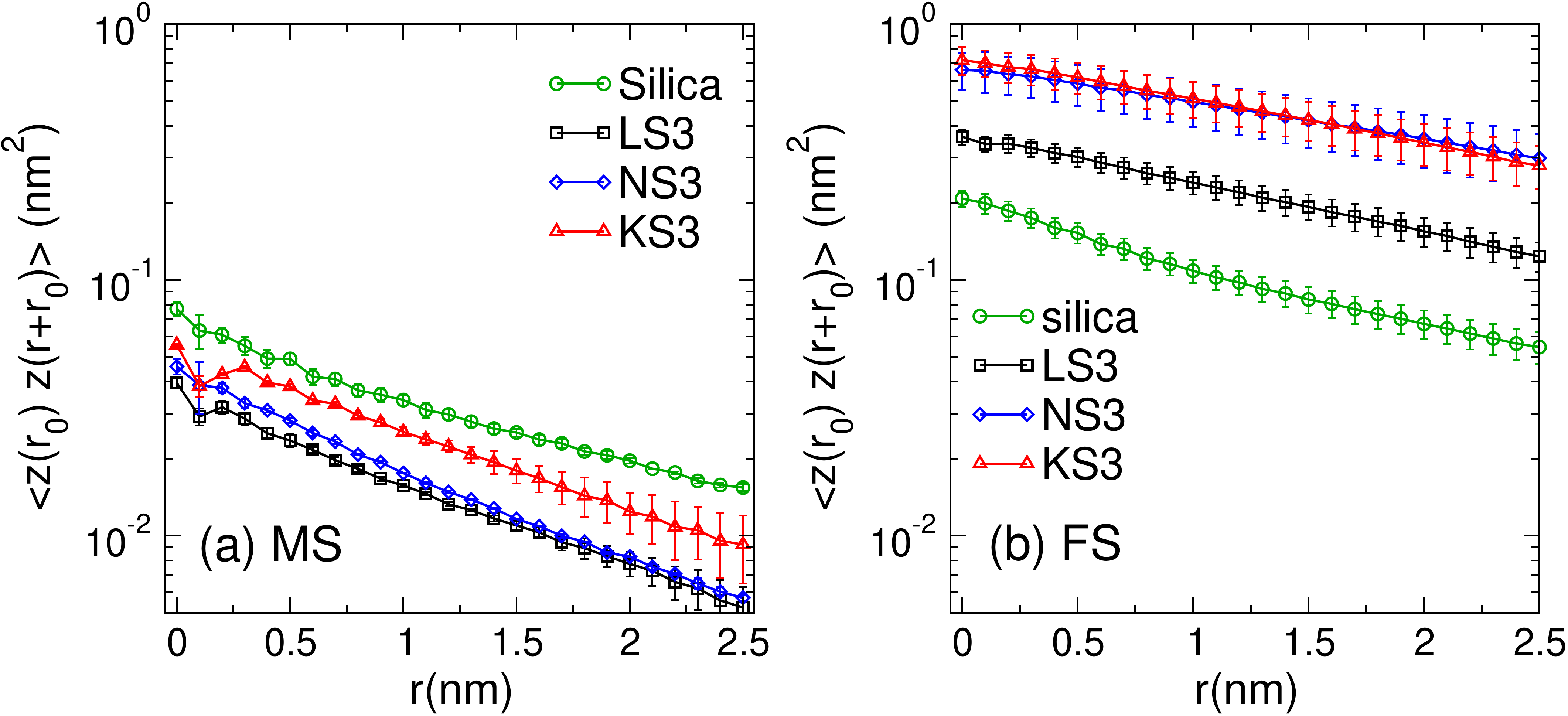}
    \caption{
    Height autocorrelation function (log-linear scale) for the MS (a) and FS (b).
    }
    \label{fig_surf-acf-new}
\end{figure}

For the FS, experimental studies have  reported a power-law dependence of $\Delta z(r)$ (no square!) on $r$ which indicates that the FS is a self-affine fractal~\cite{mandelbrot_fractal_1984,ponson_two-dimensional_2006,pallares_roughness_2018}. However, the validity of this description down to the nanometer is questionable, since at such small scales the reliability of the experimental data can be severely restricted by the spatial resolution of the measurements that use techniques such as AFM~\cite{schmittbuhl_reliability_1995,
mazeran2005curvature,lechenault_effects_2010}.   In fact, theoretical and numerical
studies of the FS of heterogenous materials~\cite{ramanathan_quasistatic_1997,
bares_nominally_2014} have reported a logarithmic dependence of $\Delta z(r)$ on $r$, but it was only 
very recently that computer simulations of realistic oxide glasses demonstrated that the FS of silica glass does indeed show such a logarithmic $r-$dependence of $\Delta z(r)$, demonstrating that on length scales less than 10~nm the surface is not a fractal object~\cite{zhang2021roughness}. In Fig.~\ref{fig_surf-scaling}(b) we show that this logarithmic dependence also holds for the FS of alkali silicate glasses and that this dependence is unaffected by varying the alkali species. We mention that this scaling property is found in both the direction of crack propagation as well as the direction parallel to the crack front (not shown). 
Finally, we note that also here the flattening of the curves at large $r$ is most likely related to the fact that the sample is finite and hence fluctuations are bounded. However, for small-to-intermediate $r$ these finite size effects are small and the observed scaling behavior is stable.

Figure~\ref{fig_surf-acf-new} shows the (not normalized) height-autocorrelation function $\langle z(r_0)\cdot z(r_0+r) \rangle$ (in contrast to the normalized one in Eq.~(\ref{eq_Cr})). By its definition, this quantity contains thus information about the surface height fluctuations and their spatial correlations. Specifically, the value of these curves at $r=0$ is given by the surface roughness, thus following the same order as $\sigma$ presented in Fig.~\ref{fig_surf-roughness}. Comparing panels~(a) and (b) demonstrates that the curves for the MS, notably the AS3 ones, decay faster than the FS curves, consistent with the difference in decay length as measured from the normalized correlation function $C(r)$ (see Fig.~\ref{fig_surf-acf}).
Furthermore one notes that on the length scales displayed in these panels, the decay can be described reasonably well by an exponential law, although we have seen in the context of Fig.~\ref{fig_surf-scaling} that this is not the case, at least not at small distances. One thus concludes that high quality data is needed in order to be able to identify reliably the $r$-dependence of the fluctuations.
Furthermore the graphs also show that the different curves do not cross each other which implies  that the relative roughness is independent of the length scale considered.
All together, Figs.~\ref{fig_surf-roughness} to~\ref{fig_surf-acf-new} present in a comprehensive and coherent manner the topographical features of the surfaces.

\section{Conclusions and outlook} \label{summary}

In this work, we have performed large scale MD simulations in order to investigate systematically
how the properties of alkali glass surfaces depend on the chemical nature of the alkali species and the surface type, i.e., surface formed by a melt-quench process (MS) or by a dynamic fracture process (FS). The main conclusions are:

\begin{enumerate}[label=(\roman*)]

\item The MS is more enriched in alkali and more depolymerized than the FS, while the elemental concentrations on the FS show a stronger dependence on the alkali type than the MS. 
\item Larger alkali ions facilitate the restoration of undercoordinated Si defects, whereas the two-membered rings created by the violent fracture process can hardly be annealed away at room temperature.
\item The presence of the surface induces a gradient in  composition in the regions beneath the surface. Larger alkali  ions lead to a stronger deviation of the surface composition from the nominal bulk composition (in particular Si concentration) in the first few atomic layers below the surface, as well as a faster decay of the compositional fluctuation. However, the zone in which the surface influences the composition extents only over 2-3~nm while beyond this distance one recovers the bulk composition to within 1~\%.
\item The roughness of the MS increases with increasing alkali size, a trend that can be attributed to the reduced surface tension, and the prediction of the theory of frozen capillary waves. The FS is found to be significantly rougher than the MS. That this roughness increases with increasing alkali size can be related to the enhanced heterogeneity in medium-range structure of the glass as well as the weakening of the bonding between the alkali atoms and the network.   
\item The decay length of surface height fluctuations of the FS (2.5~nm) is about twice the value for the MS, suggesting a more heterogeneous structure near the FS than the MS. 
 The scaling property of the MS is consistent with the prediction of the capillary wave theory, whereas we find that the FS are not self-affine fractals on the nanoscale. This conclusion is unaffected by the variation of alkali species. 

\end{enumerate}

As a final remark, we note that the surface properties of glasses are intriguing yet challenging to probe in experiments and rationalize on a microscopic level because of the disordered nature of the glass structure. The present work provides an atomistic understanding on how/why the alkali modifiers influence the surface properties of silicate glasses which is of practical relevance for the design of glasses with functional properties. In addition, the set of tools developed in this work (and our recent studies~\cite{zhang_surf-vib_2020,zhang2021roughness,zhang2021abinit}) will be useful for future investigation of the surface properties of network-forming amorphous materials.

\section{Acknowledgments}
W.K. is a senior member of the Institut Universitaire de France. The simulations were performed using the HPC resources of CINES under the allocations (Grant Nos. A0050907572 and A0070907572) attributed by GENCI (Grand Equipement National de Calcul Intensif), as well as the computing resources at Xi'an Jiaotong University.

\normalem  
\clearpage
%
\onecolumngrid

\end{document}